\documentclass{PoS}
\usepackage{slashed}
\usepackage{amsmath}
\usepackage{graphicx}
\usepackage{setspace}
\newcommand{\pa}{\partial}
\newcommand{\bphi}{\overline{\phi}}
\newcommand{\bpsi}{\overline{\psi}}

\newcommand{\vepsb}{\overline{\varepsilon}}
\newcommand{\vp}{\varphi}
\def\d{\delta}
\newtheorem{lemma}{Lemma} 
\newcommand{\lambdab}{\overline{\lambda}}
\newcommand{\arxiv}[1]{[arXiv:\href{http://arxiv.org/abs/#1}{{\tt #1}}]}
\title{Supersymmetry on the lattice and the status of the Super-Yang-Mills simulations}

\ShortTitle{Lattice SUSY and the SYM simulations}

\author{\speaker{Georg Bergner}\\
        Theoretisch-Physikalisches Institut, Universit\"at M\"unster, D-48149   M\"unster, Germany\\
        E-mail: \email{G.Bergner@uni-muenster.de}}

\abstract{Supersymmetry (SUSY) and supersymmetric field theories are an interesting topic for numerical lattice simulations.
          Similar to the chiral symmetry there is also no local realization of (interacting) supersymmetry on the lattice.
          I briefly review the basic reasons for the breaking of supersymmetry.
          One attempt to solve the problem uses a Ginsparg-Wilson relation for supersymmetry. 
          However, apart from the free theory a solution of this relation has so far not been found.
          For supersymmetric Yang-Mills (SYM) theory  a fine-tuning of the bare gluino mass is enough to arrive at a supersymmetric continuum limit.
          The last part of this work contains a short status report of recent SYM simulations.}

\FullConference{The XXVIII International Symposium on Lattice Field Theory, Lattice2010\\
		June 14-19, 2010\\
		Villasimius, Italy}

\begin{document}
\section{Introduction}
Supersymmetric lattice simulations have been the subject of several recent investigations. 
The reason for this can be seen in the growing interest for possible extensions of the standard model.
A complete understanding of supersymmetric theories, as they are used in these extensions, needs non-perturbative methods.
Supersymmetry transforms fermionic and bosonic particles into each other.
As a consequence it predicts a pairing of bosonic and fermionic states and has nontrivial commutation relations with the Poincare-symmetry of space-time.

The simplest examples for supersymmetric field theories are Wess-Zumino modells. They can be seen as the matter sector of the supersymmetric extensions of the standard model.
The continuum action of such a model has the following form\footnote{In this case the on-shell version of the ${\cal N}=2$ Wess-Zumino model in two dimensions.}
\begin{equation}
   S =\!\! \int\!\! d^2 x\,\left[\frac12
  \pa_\mu\phi\,\pa^\mu\bphi
+ \frac12 |W'(\phi)|^2
+ \bpsi (\slashed{\pa}+W''(\phi)P_+ +\overline{W}''(\bphi)P_-)\psi\right]\, .
\end{equation}
It consists of bosonic and fermionic kinetic terms and a Yukawa-type interaction term.
The supersymmetry transforms the fields $\psi$ and $\phi$ into each other. Due the nontrivial interplay with the Poincar\'e symmetry the transformations contain derivatives of the fields.
The variation of the action in the continuum is
\begin{equation}
   \d S  =   -\vepsb\!\!\int\!\!\! dt\left[ W'(\vp)(\pa_t\psi) + \psi W''(\vp)\pa_t\vp\right]
        \stackrel{\text{(1)}}{=}-\vepsb\!\!\int\!\!\! dt\pa_t\left[\psi W'(\vp)\right] \stackrel{\text{(2)}}{=}0\, .
\end{equation}
On the lattice the equality (2) can be easily satisfied using periodic or open boundary conditions.
The violation of the first equality (1) is, however, unavoidable for a local lattice theory. It is due to the breaking of the Leibniz rule on the lattice. This fact can be stated in terms of a No-Go theorem for local lattice supersymmetry.
A version of such a No-Go theorem was presented in \cite{Kato:2008fr}. As shown in \cite{Bergner:2009vg}, even stronger conditions for a local lattice action must be fullfilled. The findings in \cite{Bergner:2009vg}, that include also the results of a simulation with intact supersymmetry on the lattice, will be summarized in the next section. A No-Go theorem for a local realization of a supersymmetry seems to be similar to the Nielson-Ninomiya theorem \cite{Nielsen:1981hk} for chiral symmetry. In this case the solution was found in terms of a modified symmetry relation on the lattice.
The Section \ref{sec:GW} is therefore a briefly comment on the current investigations of this approach.
The last Section \ref{sec:SYM}, consists of a report of the simulations of a supersymmetric gauge theory. In this case the unavoidable breaking of supersymmetry can be controlled via fine tuning.

\section{No-Go theorem for local lattice supersymmetry}
A simple form of a No-Go theorem for the Leibniz rule, and therefore for the invariance of the lattice action can be found in terms of the following statement.
\begin{lemma}[Simple form of the No-Go theorem]
For all lattice derivative operators $\nabla_{nm}$ the Leibniz-rule is violated, 
\begin{equation}
 \sum_{m} \nabla_{nm}(f_m g_m)- f_n\sum_{m} \nabla_{nm} g_m-g_n\sum_{m} \nabla_{nm} f_m\neq 0\, .
\end{equation}
\end{lemma} 
Note that this is true also for a nonlocal lattice derivative. It is therefore a stronger restriction then in \cite{Kato:2008fr}.

Already in the early days of lattice supersymmetry a generalization of the discretization was proposed in \cite{Dondi:1976tx}, that circumvents this simple form of the No-Go theorem using a modification of the product rule on the lattice. It leads, e.~g., to the following relplacement for a product of three continuum fields on the lattice
\begin{equation}
 \int dx\phi(x)^3 \rightarrow \sum_{i,j,k} C_{ijk} \phi_i\phi_j\phi_k\, .
\end{equation}
 However, the lattice action in \cite{Dondi:1976tx} includes a nonlocal $C_{ijk}$. The nonlocality of $C_{ijk}$ is indeed unavoidable to circumvent the simple form of the No-Go theorem. Furthermore, $C_{ijk}$ projects one of the interacting fields to its zero momentum part and can hence be considered only as a trivial solution.
To exclude such kind of trivial solutions one has to require an additional condition for a possible lattice action. 
The interaction should not be restricted to a only a discrete number of modes.

A realization that fullfills this condition has been proposed in \cite{Bartels:1983wm}. In contrast to the previously mentioned approach it involves not only a nonlocal interaction term but also a nonlocal SLAC type derivative.\footnote{For a definition of this derivative see \cite{Drell:1976mj}.} 
This seems to be a severe violation of locality, but for a supersymmetric lattice theory one can find no better solution. This can be stated in terms of the following No-Go statement. 
\begin{lemma}[No-Go theorem]
In order to get a nontrivial interacting supersymmetric lattice theory one needs a nonlocal derivative operator and a nonlocal interaction term.
\end{lemma}
Besides the violation of the Leibniz rule a second source of the supersymmetry breaking on the lattice must be mentioned. Due to the doubling problem in the fermionic sector a Wilson mass term must be added to the fermion action. If not consistently added also to the bosonic potential this leads to a supersymmetry breaking.
Note that the doubling problem can also be avoided with a nonlocal lattice action.

\begin{figure}
\centering
\includegraphics[width=10cm]{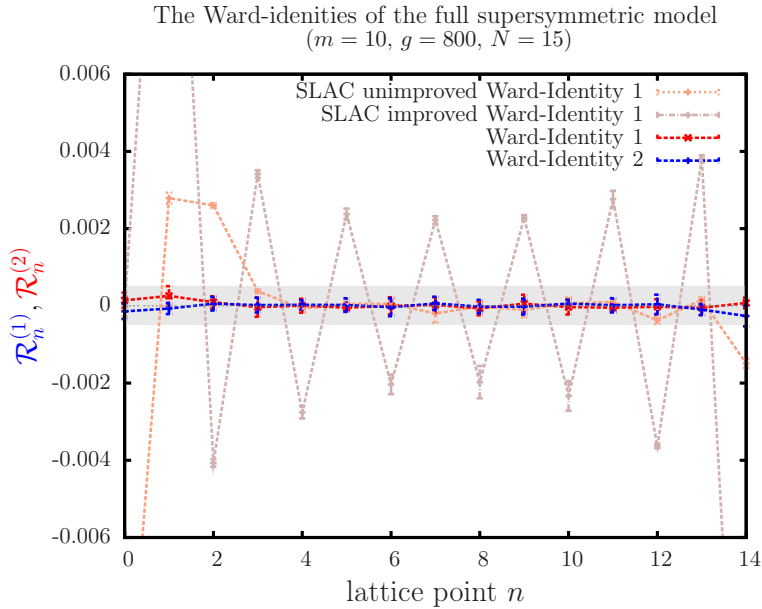}
\caption{The Ward-identities measured in supersymmetric quantum mechanics using a model with completely realized supersymmetry on the lattice. A nonlocal SLAC derivative and a nonlocal interaction term were used in the action. The figure shows for comparison also the Ward-identities of the unimproved and improved SLAC model introduced in \cite{Bergner:2007pu}. These models contain only the nonlocal SLAC derivative but no nonlocal interaction term.}
\label{fig:WISQM}
\end{figure}

According to the No-Go theorem it is hence possible to realize a fully supersymmetric theory on the lattice, if one accepts a nonlocality of the action. A local continuum limit of such a theory can be proven to all orders of perturbation theory in Wess-Zumino models up to three dimensions \cite{Kadoh:2009sp}.
To perform numerical simulations one has to seek for an efficient realization of the nonlocal interaction term on the lattice. This can be done by performing the complete simulation in Fourier space. Equivalently the nonlocal interaction term can be realized using Fourier transformations and a local product on a larger lattice (cf.\ \cite{Bergner:2009vg} for details). Figure \ref{fig:WISQM} shows the result for the supersymmetric Ward-Identities, that verify the intact supersymmetry on the lattice.
\section{Solutions similar to the Ginsparg-Wilson relation}
\label{sec:GW}
Considering the presented No-Go theorem, the problem of realizing supersymmetry on the lattice seems to be much similar to the one for chiral symmetry on the lattice.
The corresponding No-Go theorem for local realizations of chiral symmetry is the Nielsen-Ninomiya theorem. In case of this symmetry a solution for a lattice realization was found in terms of the Ginsparg-Wilson relation, a modified symmetry relation on the lattice \cite{Ginsparg:1981bj}.
The basic idea behind the Ginsparg-Wilson relation is a blocking transformation (corresponding to a renormalization group step) from the continuum to the lattice. In this way not only the action is mapped onto a (perfect) lattice action. Also the symmetry transformations get a modification due to the symmetry breaking induced by the regulator.
In case of chiral symmetry the relevant part of the action is only quadratic and a solution for the modified symmetry transformation can be found easily.
In contrast, the interesting case of supersymmetry includes non-quadratic interaction terms. A modified symmetry relation can, nevertheless, be found, as presented in \cite{Bergner:2008ws}. 
It is, however, hard to find an action, that is invariant under it.
The form of the modified symmetry leads generically to non-polynomial actions and involves possible nonlocal terms. 
This might not be unexpected since a renormalization group transformation generically generates operators of higher order in the fields and interactions between distant lattice sites.
In case of the Ginsparg-Wilson relation a solution for a lattice action invariant under the modified chiral symmetry was found, e.~g., in terms of the overlap operator. For supersymmetry this problem remains unresolved.

A different way to ensure chiral symmetry on the lattice uses fine-tuning of the parameters in the action. The number of parameters that must be fine tuned depends on the mixing of the supersymmetry breaking terms with operators of equal or lower dimension. Based on similar arguments one can find that no fine tuning is needed for the Poincar\'e symmetry and only a fine tuning of the mass parameter for the chiral symmetry \cite{Bochicchio:1985xa}.
A fine tuning of all necessary parameters for a Wess-Zumino model (in more than one dimension) seems currently impossible. However, for a supersymmetric Yang-Mills theory it was found that only one parameter is needed for the fine tuning \cite{Curci:1986sm}.
\section{Supersymmetric Yang-Mills theory}
\label{sec:SYM}
The field content of the supersymmetric ($\mathcal{N}=1$) Yang-Mills theory contains, besides the usual bosonic gauge fields (field strength $F_{\mu\nu}$), the Majorana fermion $\lambda$  in the adjoint representation.
Adding a gluino mass term the Lagrangian of the continuum theory has the following form
\begin{equation}
 \mathcal{L}=\text{Tr}\left[-\frac{1}{4} F_{\mu\nu}F^{\mu\nu}+\frac{i}{2}\lambdab\slashed{\mathcal{D}}\lambda-\frac{m_g}{2}\lambdab\lambda\right]\, .
\end{equation}
Note that supersymmetry is only established at $m_g=0$.
Low energy effective actions of the theory have been constructed in \cite{Veneziano:1982ah,Farrar:1997fn}. 

The lattice action proposed by Veneziano-Curci in seems, in comparison to the previously mentioned attempts, to be a  rather ``brute force'' approach. It constits of the usual gauge action (gauge group SU($N_C$)) and the common fermionic action with a Wilson term,\footnote{$U_P$ denotes the usual plaquette and $D_w$ the Dirac matrix including the gauge field, the Wilson term, and the mass term of $m_g$.}
\begin{equation}
 \mathcal{S}_L=\beta \sum_P\left(1-\frac{1}{N_c}\Re U_P\right) +\frac{1}{2}\sum_{xy} \lambdab_x\left( D_w(m_g)\right)_{xy}\lambda_y\, .
\end{equation}
It breaks the chiral symmetry and the supersymmetry of the model. However, a detailed analysis shows that both symmetries can be recovered with a fine-tuning of the bare gluino mass $m_g$. The chiral limit of the theory corresponds to the supersymmetric limit.

We applied this approach to perform numerical simulations using a PHMC algorithm. For the gauge action (gauge group SU(2)) an additional tree-level Symanzik improvement was used and stout smeared links for the fermion action.
We have considered lattice sizes of $16^3\times 32$, $24^3\times 48$, and $32^3\times 64$ lattice points.
If one sets the Sommer scale of the theory to the usual QCD value ($r_0=0.5\,\text{fm}$) the lattice spacing is $0.09\,\text{fm}$ corresponding to a lattice volume of $L\approx 1.5-2.3 \,\text{fm}$.
We have performed the fine tuning by an extrapolation to the chiral limit.\footnote{The mass of the connected part of the  $\lambdab \gamma_5 \lambda$ correlation vanishes at that point. It corresponds to the adjoint version of a pion mass ($m_{a-\pi}$) in a partially quenched framework} It was confirmed that the supersymmetric Ward identities would lead to the same extrapolated point.\footnote{The renormalized masses obtained form the chiral and the supersymmetric Ward identities vanish at the same point. Up to $O(a)$ effects the breaking of the corresponding symmetry is determined by these masses.}
Although we have used anti-periodic boundary conditions for the fermions in time direction, the finite volume effects, including the supersymmetry breaking by the boundary conditions, seem to be under control.

The proposed low energy effective action contains operators for adjoint mesons (gluino-balls), glueballs, and, since the fermions are in an adjoint representation, compound operators formed from gluon and gluino fields (gluino-glue-balls).
It is a nontrivial task to measure the corresponding correlations on the lattice. All mesonic states contain disconnected contributions. We have measured them with a stochastic estimator technique including, where necessary, a separate determination of the contributions form the lowest eigenvalues.
For the glueballs we applied variational smearing methods, and the gluino-glue-ball correlation was measured with a combination of Jacobi and APE smearing. 

The presence of Majorana fermions in the theory leads to additional difficulties. 
Instead of the determinant the fermionic path integral corresponds to a Pfaffian of $D_w$. Up to a sign the Pfaffian is the square root of the determinant. Hence one gets a sign problem in the theory, although the determinant stays always positive.
The signs are represented in our approach by positive and negative reweighting factors.

\begin{figure}
\centering
 \includegraphics[width=10cm]{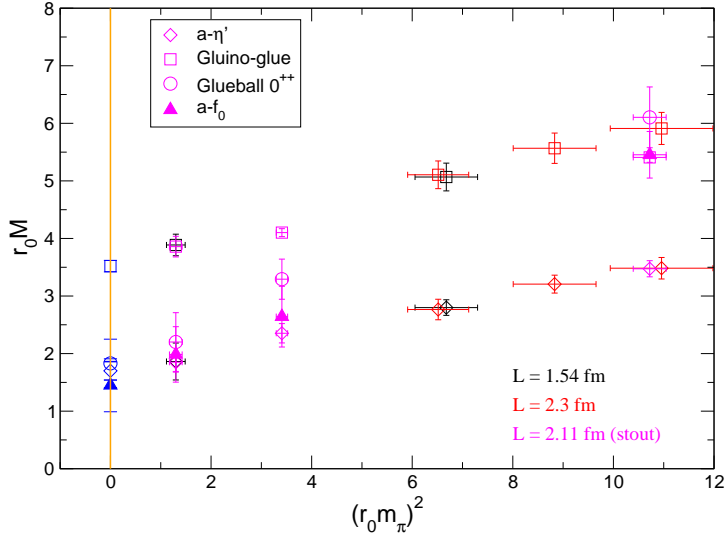} 
\caption{The masses of the particles obtained in our simulation of supersymmetric (SU(2)) gauge theory as a function of the adjoint pion mass. The blue symbols correspond to the values extrapolated to the chiral limit. All masses are given in units of the Sommer scale $r_0$.}
\label{fig:SpectrumSUSYM}
\end{figure}
The masses of the particles obtained from the correlation functions are shown in figure \ref{fig:SpectrumSUSYM}.
The gluino-glue-ball operator $\text{Tr}_c[F\sigma\lambda]$\footnote{$F_{\mu\nu}(x)$ is replaced by the clover plaquette on the lattice, and $\sigma^{\mu\nu}$ is the commutator of two gamma matrices.} should have an overlap with the lightest fermionic state of the theory. If supersymmetry is not broken, it should be paired with bosonic particles of the same mass.
We have obtained masses for the gluino-ball $a$-$f_0$ (operator $\lambdab\lambda$) and $a$-$\eta^\prime$ (operator $\lambdab\gamma_5\lambda$) as well as the $0^{++}$ glueball.
However, all of these particles have a much smaller mass than the gluino, when extrapolated to the chiral limit. This is in contradiction with the theoretical predictions since it indicates a breaking of supersymmetry. 

In our recent studies we have increased the value of $\beta$ to approach a smaller lattice spacing (around $0.06 \,\text{fm}$).
The first results of these simulations indicate that the gap between the masses of the fermionic and bosonic masses is reduced towards the continuum limit.
However, it is still too early to draw a conclusion from these preliminary data.
\section{Conclusions}
Supersymmetry on the lattice remains an interesting subject of theoretical investigations. 
It is by now well understood that this symmetry can be realized only in a nonlocal theory on the lattice.
It has been shown in \cite{Karsten1981103} that a nonlocal lattice gauge theory leads to a nonlocal continuum limit.
However, in case of the Wess-Zumino modells a full supersymmetric theory yields the correct local continuum limit at least in lower dimensions.
 
In a local lattice theory it is necessary to control the breaking of supersymmetry.
The most elegant way to control the breaking would be a Ginsparg-Wilson relation for supersymmetry.
Although it is possible to establish such an relation, it is hard to find a solution in the interacting case.
Perhaps an approximation can lead to a useful lattice realization \cite{BBPISE}.

For the supersymmetric counterpart of a pure gauge theory the supersymmetry breaking can be controlled with the same fine tuning as needed for the chiral symmetry.
Only a single parameter is used to arrive at a supersymmetric continuum limit.
This theoretical prediction is verified by the supersymmetric Ward-Identities.
The investigation of the mass spectrum of the theory still demands further efforts. The first results at one lattice spacing show a gap between the bosonic and fermionic masses. This gap may, however, be due to the lattice artifacts and vanish in the continuum limit.
\section*{Acknowledgments}
The results of the simulations of supersymmetric Yang-Mills theory have been obtained in collaboration with K.~Demmouche, F.~Farcioni, G.~M\"unster, I.~Montvay and J.~Wuilloud. I thank G.~M\"unster for corrections and useful remarks.
\begin{spacing}{0.9}

\end{spacing}
\end{document}